\documentclass[aps,pra,twocolumn,showpacs,superscriptaddress,floatfix]{revtex4-1}
\usepackage{graphicx,amsmath,amssymb}
\usepackage{bm}
\usepackage[usenames]{color}
\usepackage[dvipsnames]{xcolor}
\usepackage[unicode=true,pdfusetitle,
 bookmarks=false,
 breaklinks=false,pdfborder={0 0 1},backref=false,colorlinks,urlcolor=Black, linkcolor=blue,citecolor=blue]
 {hyperref}
\begin{document}

\title{Exotic Quantum States in Spin-1 Bose-Einstein Condensate with Spin-Orbit Coupling in Concentric Annular Traps
}

\author{Yun Liu}
\affiliation{School of Physical Science and Technology, Lanzhou University, Lanzhou 730000, China}
\affiliation{Key Laboratory for Quantum Theory and Applications of MoE, Lanzhou Center for Theoretical Physics, Lanzhou University, Lanzhou 730000, China}

\author{Zu-Jian Ying}
\email{yingzj@lzu.edu.cn}
\affiliation{School of Physical Science and Technology, Lanzhou University, Lanzhou 730000, China}
\affiliation{Key Laboratory for Quantum Theory and Applications of MoE, Lanzhou Center for Theoretical Physics, Lanzhou University, Lanzhou 730000, China}

\begin{abstract}
We explore the exotic quantum states emerging in the ground state (GS) of a strongly-correlated spin-1 Bose-Einstein condensate confined in two-dimensional concentric annular traps with a spin-orbit coupling (SOC). In the antiferromagnetic case, the GS density manifests various patterns of distributions, including facial-makeup states, petal states, topological fissure states, multiple-half-ring states and  property-distinguished vertical and horizonal stripe states. We notice a peculiar phenomenon of density-phase separation in the sense that the variations of density and phase tend to be independent. In ferromagnetic case, the GS exhibits a semi-circular or half-disk status of density embedded with vortices and anti-vortices. The spin distribution can self-arrange into an array of half-skyrmions and we also find a half-antiskyrmion fence separating vortex-antivortex pairs. Our study indicates that one can manipulate the emergence of exotic quantum states via the interplay of the SOC, interaction and potential geometry and the abundant state variations might also provide potential resources for quantum metrology.
\end{abstract}

\maketitle

\section{Introduction}

Spin-orbit coupling (SOC) has a wide relevance in physical systems such as condensed matter~\cite{Krieger2024parallelSOC,JIANG2017PhysRepSkyrmions,ManchonNatMat-Rashba2015,Rashba1984,Dresselhaus1955},
nanosystems~\cite{Gentile2022NatElec,Ying2020PRR,Ying2017curvedSC,Ying2016Ellipse,Nagasawa2013Rings}, light-matter interactions~\cite{Ying-2021-AQT,Ying-gapped-top,Ying-Stark-top,Ying-Spin-Winding,Ying-JC-winding} and cold atoms~\cite{ref33,Liu2024oscillatory,chen2024solitonSOC,GaoXianlong2023BECsolitons,GongMing2019BECSOC,LinRashbaBECExp2013Review,Li2012PRL,LinRashbaBECExp2011}. The competition and interplay of SOC with spin interaction, density interaction, external potential, and geometry may induce exotic quantum states and novel quantum phases of matter.

Among these relevant systems a highly-controllable system is ultra-cold atomic gas~\cite{Li2012PRL,LinRashbaBECExp2011,LinRashbaBECExp2013Review}. The alkali metal atomic gases in the optical trap are bound to hyperfine energy states, forming a spinor Bose-Einstein condensate (BEC)~\cite{ref1,ref2}. Spinor BEC provides more opportunities to study various interesting topological excitations due to the release of internal spin degrees of freedom~\cite{ref3,ref4,ref5,ref6,ref7,ref8,ref9,ref10}. Through the application of the Feshbach resonance technology, one can continuously adjust the s-wave scattering length between atoms to accurately control the internal and external degrees of freedom of atoms~\cite{ref11,ref12,ref13}. Such a high controllability of BEC provides an ideal platform to explore novel physical properties and exotic quantum states~\cite{ref14,ref15,ref16}.

In particular, synthetic spin-orbit coupling (SOC) can be realized in ultracold atomic spinor BEC systems~\cite{LinRashbaBECExp2011,LinRashbaBECExp2013Review,wu2016realizeSOC,Campbell2016RealizeSpin1SOC}. SOC plays a crucial role in many important condensed matter phenomena~\cite{ref17,ref18,ref19}. Still, more varieties of properties stem from the interplay of SOC with density and spin interactions as well as the geometry and dimensions of the external potential, including plane-wave phase and stripe phase~\cite{ref19,GongMing2019BECSOC}, solitons~\cite{GaoXianlong2023BECsolitons,ref21,ref22,ref23,ref24}, vortices~\cite{ref25,ref26,ref27}, supersolids~\cite{ref28,ref29}, lattice phases~\cite{ref30}, skyrmions~\cite{ref31} and so forth. Regarding the shape of the external potential, different types of potential can be considered within the current experimental capability, such as the harmonic trap~\cite{ref32,ref33,ref34}, optical lattice~\cite{ref35,ref36,ref37}, toroidal trap~\cite{ref38,ref39,ref40,ref41,ref42,ref43}, double wells~\cite{ref44}, vertically or concentrically coupled double-ring traps~\cite{ref45}.

When most studies have been concentrated on the conventional SOCs including Rashba and Dresselhaus SOCs~\cite{Liu2024oscillatory,chen2024solitonSOC,MengPRL2016GasSOC,Li2012PRL,ref33,wu2016realizeSOC,Campbell2016RealizeSpin1SOC,LinRashbaBECExp2013Review,Anderson2013PRLmagnGenerateSOC,LinRashbaBECExp2011}, an unconventional SOC is introduced recently~\cite{ref33} for spin-1 BEC in a harmonic potential, which induces a novel topological structure as vortex molecule. A deeper study is desirable for the interplay of the unconventional SOC with the density/spin interactions and the potential geometry. Especially, one may expect that more exotic quantum states may arise when the interplay situation is affected by the potential geometry.

In the present work we consider a spin-1 BEC with the unconventional SOC in concentric annular traps which spatially creates a more stimulating situation for the interplay of the SOC with the density interaction and spin interaction. We find that such a change of potential geometry induces abundant exotic quantum states in the ground state (GS). Indeed, in the antiferromagnetic case, the GS density manifests various patterns of distributions, including petal states, facial-makeup states, topological fissure states, multiple-half-ring states and property-distinguished vertical and horizontal stripe states. In a spin-1 component we notice a peculiar phenomenon of density-phase separation. In ferromagnetic case, the GS exhibits semi-circular and half-disk states embedded with vortices and anti-vortices. We also find states with a self-arranging array of half-skyrmions and half-antiskyrmion fence separating vortex-antivortex pairs.

This paper is organized as follows.
In Sec.~\ref{Model}, the model of the spin-1 BEC with the unconventional SOC confined in concentric annular traps is introduced, and the numerical method is briefly given.
In Sec.~\ref{Sect-AntiF} we address the GS structure in antiferromagnetic case. The effects of SOC and the influence of the interaction ratio are analyzed, various patterns of distributions are demonstrated and the phenomenon of density-phase separation revealed.
In Sec.~\ref{Sect-FM} we discuss the GS in ferromagnetic case. Density distribution style different from the antiferromagnetic case are demonstrated. In particular, topological defects, including vortices and anti-vortices, array of half-skyrmions and half-antiskyrmion fence, are unveiled.
Finally, Sec.~\ref{Conclusion} gives a summary of the main results.

\section{Model and method}
\label{Model}

We consider a quasi-two-dimensional spin-1 BEC with the unconventional SOC confined in concentrically coupled annular traps. Within the mean-field approximation, the effective Hamiltonian can be written as $\mathcal{H}=\mathcal{H}_{0}+\mathcal{H}_{int}$, with
\begin{align}
	 &\mathcal{H}_{0} = \int d \boldsymbol{r} \biggr[\Psi^{\dagger}\Big(-\frac{{\hbar}^{2}{\nabla}^{2}}{2m} + V(\boldsymbol{r}) + \nu_{so} \Big)\Psi \notag \biggr], \\
	 &\mathcal{H}_{int} = \int d\boldsymbol{r} \biggr[\Psi^{\dagger}\Big(\frac{c_{0}}{2} {n}^2 + \frac{c_{2}}{2}|\boldsymbol{F}|^2\Big) \Psi  \biggr],
\end{align}
where $\Psi=(\psi_{1},\psi_{0},\psi_{-1})^{T}$ denotes the order parameter of BEC, subject to the normalization condition $\int d\boldsymbol{r}\Psi^{\dagger}\Psi=N$, with $N$ being the total particle number. $m$ is the mass of atom and $n=n_1+n_0+n_{-1}=\sum_{m_F}\psi^{\dagger}_{m_F}\psi_{m_F}$, where $m_F=0,\pm 1$, is the total condensate density of all spin components. $V(\boldsymbol{r})$ is the external potential, which will be given below. Here the effective space is restricted to the $x$ and $y$ dimensions, with the position denoted by $\boldsymbol{r}=(x,y)$, while the dynamics in the z-direction is frozen by a strong confinement. The unconventional SOC with coupling strength $\kappa$ takes the form of  $\nu_{so}=-i \hbar \kappa(f_z \partial_x+f_y \partial_y)$~\cite{ref33}, with $\boldsymbol{f}=(f_{x},f_{y},f_{z})$ being the
vector of the spin-1 matrices. The unconventional SOC contains both orthogonal~\cite{Rashba1984} and parallel~\cite{Krieger2024parallelSOC} spin-momentum couplings and may be synthesized in ultracold atomic spinor BEC systems~\cite{ref33,LinRashbaBECExp2011,LinRashbaBECExp2013Review,wu2016realizeSOC,Campbell2016RealizeSpin1SOC}.
The coefficients $c_{0}$ and $c_{2}$ in $\mathcal{H}_{int}$ represent the strengths of density-density and spin-exchange interactions respectively, while $\boldsymbol{F}=\Psi^{\dagger}f_{\alpha}\Psi(\alpha=x,y,z)$ is the spin density vector.

The condensate can be described by the Gross-Pitaevskii equation
\begin{equation}
i\hbar \frac{\partial }{\partial t}\Psi (\boldsymbol{r},t)=\left\{ h(
\boldsymbol{r},t)+n(\boldsymbol{r},t)\left[ c_{0}+c_{2}\boldsymbol{F}(
\boldsymbol{r},t)\cdot \boldsymbol{f}\right] \right\} \Psi (\boldsymbol{r},t)
\end{equation}
where $h(\boldsymbol{r},t)=-\hbar ^{2}\nabla ^{2}/(2m)+V(\boldsymbol{r}
)+\kappa \boldsymbol{f}\cdot \boldsymbol{p}$ and $\boldsymbol{p}=(0,-i\hbar
\partial _{y},-i\hbar \partial _{x})$.
In order to facilitate numerical treatments, the units for energy, length, time, and $\kappa$ are rescaled by $\hbar\omega_{r}$, $\sqrt{\hbar/ m\omega_{r}}$, $1/\omega_{r}$ and $\sqrt{\hbar \omega_{r} /m}$, respectively. With these units, the coefficients become $c_0=4N\sqrt{m\pi\omega_z/2\hbar}(2a_{2}+a_{0})/3$ and $c_2=4N\sqrt{m\pi\omega_z/2\hbar}(a_{2}-a_{0})/3$, where $a_0$ and $a_2$ are the s-wave scattering lengths in the total spin channels. The equation becomes dimensionless by setting $m=\hbar=1$.

\begin{figure}
    \includegraphics[width=1.0\columnwidth]{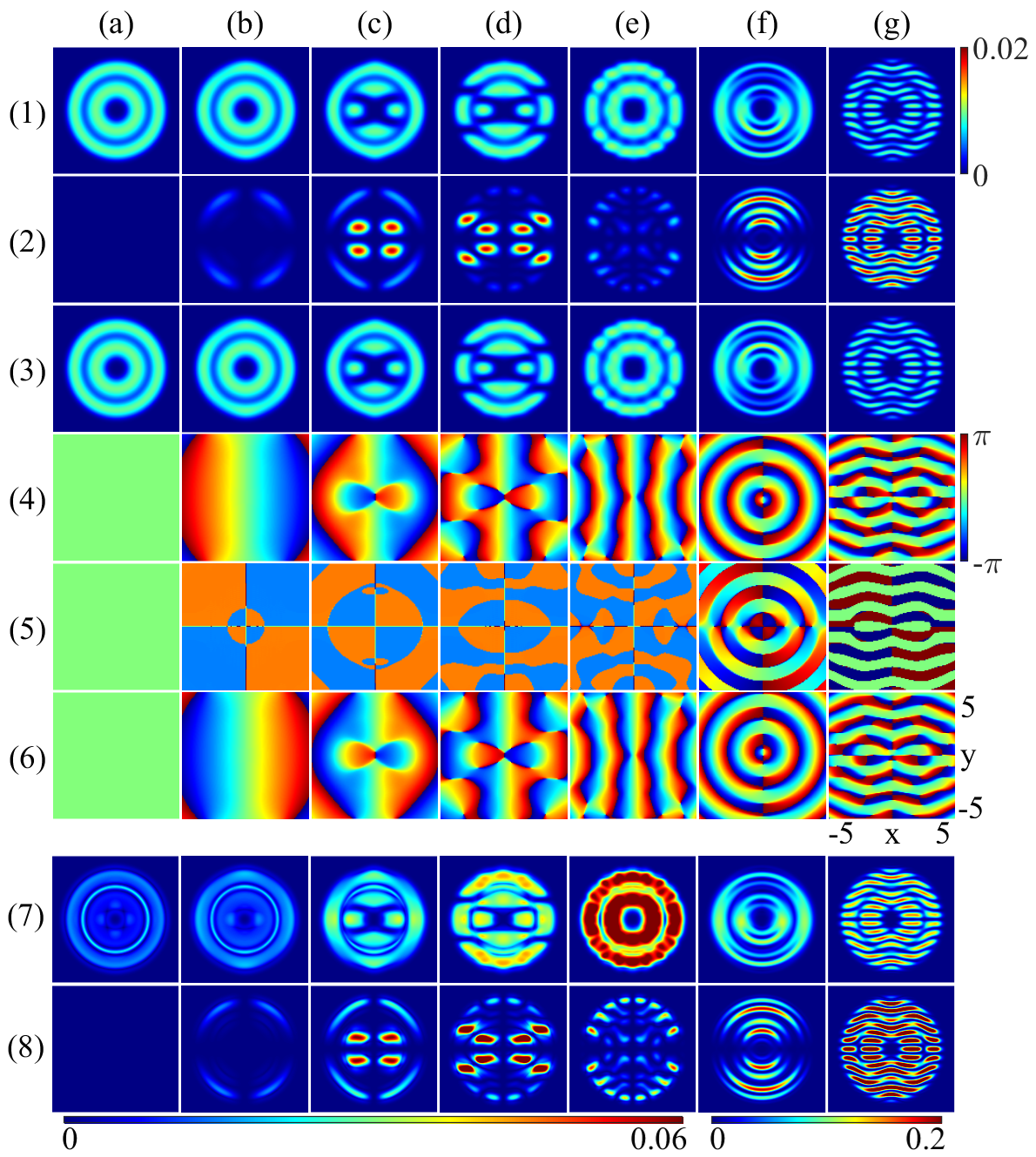}
	\caption{Distributions of density (1-3), phase (4-6) and local kinetic energy (7,8) for ground states with the antiferromagnetic interaction $c_0=550$ and $c_2=55$ in different spin-orbit coupling (SOC) strength $\kappa$. (a) $\kappa=0$, (b) $\kappa=0.6$, (c) $\kappa=1.2$, (d) $\kappa=1.6$, (e) $\kappa=2.5$, (f) $\kappa=3$, (g) $\kappa=4$; (1,4,7) $m_{F}=+1$, (2,5,8) $m_{F}=0$, (3,6) $m_{F}=-1$.}
	\label{fig-1}
\end{figure}

We consider the external trapping potential
\begin{align}
	V(r)=min\left\{\frac{1}{2}m\omega_{0}^{2}(r-R_{0})^{2},\frac{1}{2}m\omega_{1}^{2}(r-R_{1})^{2}\right\},
\end{align}
with $r$ being the distance from the origin as in the polar coordinates. The two overlapping parabolas with respect to $r$, characterized by frequencies $\omega_{0}$ and $\omega_{1}$ and the bottom locations $r=R_0$ and $R_1$, finally form two concentric annular traps in the $x$-$y$ plane. Furthermore, to ensure that the product of the 'width' and the radius of each annulus remains comparable, one can set the condition $\omega_{1} >\omega_{0}$. In this paper, we take $R_0=2a_0$ and $R_1=4a_0$ with $a_0=\sqrt{\hbar/(m\omega)}$ being the oscillator length, $\omega =\omega_{0}/4$ and $\omega_0/ \omega_{1} =4/5$.
To obtain the GS of the system, the Gross-Pitaevskii energy, given by
\begin{align}
E=&\frac{1}{2}\int d\boldsymbol{r}\sum_{m_F}|\nabla\psi_{m_F}|^{2}+2V(r)n+c_{0}n^{2} \notag \\
&+c_{2}[(n_1-n_{-1})^{2}+2|\psi_{1}^{*}\psi_{0}+\psi_{0}^{*}\psi_{-1}|^{2}] \notag   \\
&+\sqrt{2}\kappa[\sqrt{2}i({\psi}_{-1}^{*}{\partial}_{x}{\psi}_{-1}-{\psi}_{1}^{*}{\partial}_{x}{\psi}_{1}) \notag \\
&+\psi_{0}^{\ast}(\partial_{y}\psi_{1}-\partial_{y}\psi_{-1})-\psi_{1}^{\ast}\partial_{y}\psi_{0}+\psi_{-1}^{\ast}\partial_{y}\psi_{0}]
\label{E}
\end{align}
as a functional of $\Psi$, is minimized with the aid of the imaginary-time method.

\begin{figure}
    \includegraphics[width=1.0\columnwidth]{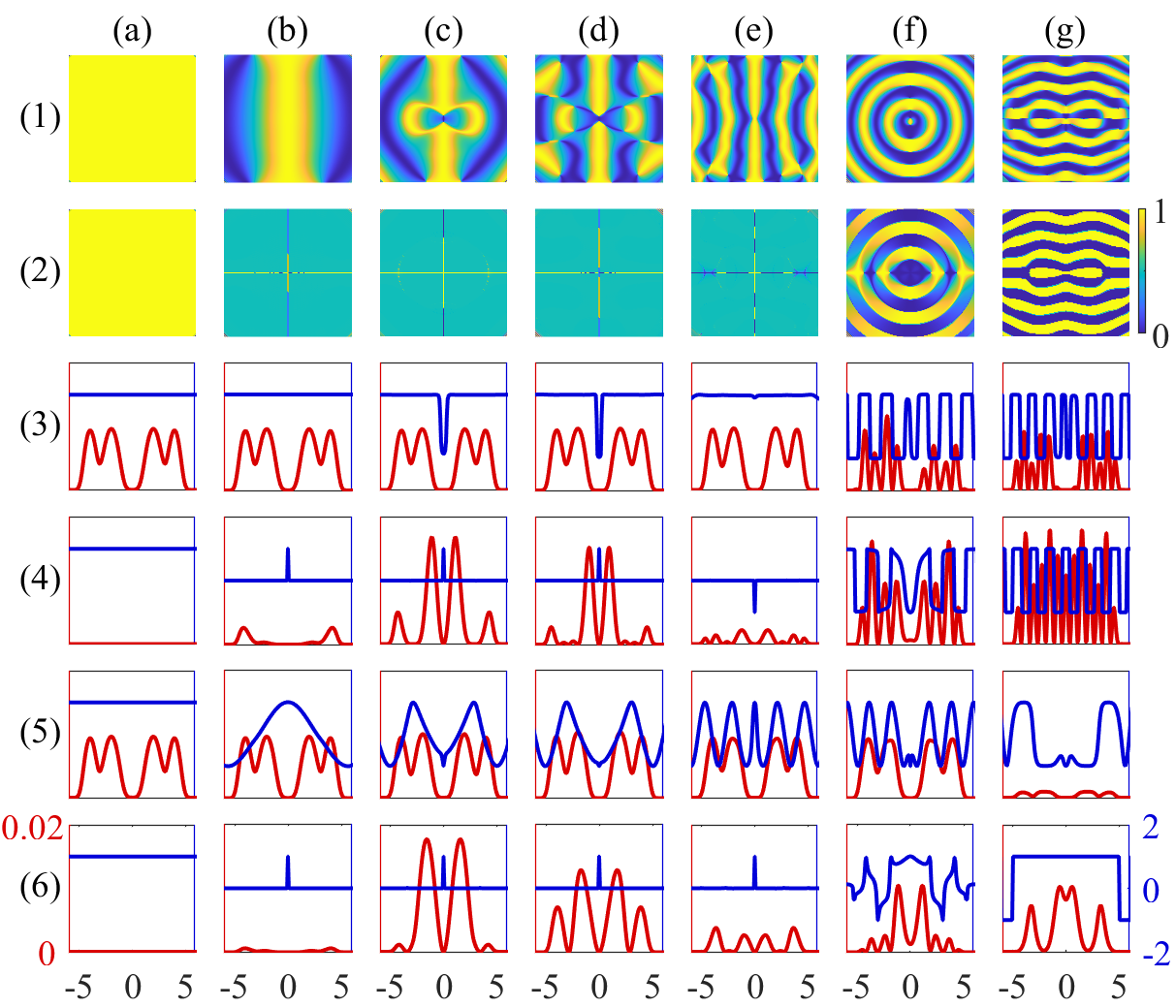}
	\caption{Density-phase separation. (1,2) Cosine plot of the phase for $m_F=+1$ (1) and $m_F=0$ (2) corresponding to the cases (a)-(g) in Fig.~\ref{fig-1}. (3-6) Line plot for the cosine phase (blue) compared with the density (red) at fixed $x=0.08$ for $m_F=+1$ (3), $x=1.25$ for $m_F=0$ (4), $y=0.08$ for $m_F=+1$ (5), and $y=1.25$ for $m_F=0$ (6). }
	\label{fig-1-phase}
\end{figure}

\section{Antiferromagnetic Case}
\label{Sect-AntiF}

In this part, we consider the GS structure in
antiferromagnetic case with $c_{2}>0$.  In various patterns of density
distributions, some exotic quantum states emerge, including facial-makeup
states, topological fissure states, property-distinguished vertical and
horizontal stripe states. A peculiar tendency of density-phase separation is
found in the sense that the variations of density and phase tend to be
independent.

\subsection{Density distributions and exotic quantum states}

Fig.~\ref{fig-1} shows the GS in varying the
strength of SOC from zero (a) to a large one (g) at fixed interactions
$(c_{0}=550,c_{2}=55)$. Here in Fig.~\ref{fig-1} the upper three rows (1-3) display the density
distributions for the spin components $m_{F}=1$ (1), $m_{F}=0$ (2), $m_{F}=-1$
(3), the middle three rows (4-6) show the corresponding phase ($\phi
_{m_{F}}$) of $\psi _{m_{F}}$ and the lower two rows (7,8) represent the
local kinetic energy $T_{m_{F}}=\psi _{m_{F}}^{\ast }\frac{-{\hbar }^{2}{
\nabla }^{2}}{2m}\psi _{m_{F}}/(\psi _{m_{F}}^{\ast }\psi _{m_{F}})$ for
$m_{F}=1$ (7) and $m_{F}=0$ (8).

In the density distributions in Fig.~\ref{fig-1}(1-3), with the increasing SOC the $m_{F}=\pm 1$ components
change from inner and outer rings (a), deformed
outer ring (b), broken inner ring (c), both broken rings (d), reconnected
inner ring and more fragmented outer ring (e), rings with fissures (f), to
rings broken into horizontal stripes (g).
In particular, the fissures in the upper and lower parts
of both rings in case (f) actually change the topological structure of the rings, just like the number of holes in an object is changed.
But the topological scenario is not only this simple geometric topology in the density.
In fact, we further find that such density-fissure state is accompanied with spin vortices or antivortices,
we leave the discussion in Section~\ref{Sect-Vortx-antiVortx} together with the ferromagnetic case.
Thus, the fissures actually change the topology both in the density and spin distributions.

The density distribution of $m_{F}=0$ [row (2) in Fig.~\ref{fig-1}] starts
with an empty filling (a), then appears as broken outer ring (b), petals in
inner part with broken outer ring (c), extended petals to outer part and
more fragmented outer ring (d), two size sets of inner petals overlapping
with outer fragmented outer ring (e), multiple half rings (f) and stripe
state (g). The vanishing distribution in (a) minimizes the spin-spin
interaction energy as the $c_{2}$ term in Eq.~\eqref{E} is positive. The half
rings in (f) fill in the density fissures of $m_{F}=\pm 1$ components to
reduce the energy of density-density interaction. And the stripes in (d) are
dislocated from those in $m_{F}=\pm 1$ components to minimize the
energy of density-density interaction as well.

Coincidentally, the broken ring
structures in the $m_{F}=\pm 1$ components appear like a human face in Balaclava (c) and a head shot of a
person with headphones (d). The mixed-petal structure of case (e) in the $m_{F}=0$ component
also looks like a facial-makeup of Peking opera [clearer view in row (8)].
Such coincidences also occur in lighter-matter interactions where the effective SOC produces spin-winding profiles in formal or spiritual similarity with humans or animals~\cite{Ying-Spin-Winding}, reflecting another interesting side of SOC in quantum effects.

We also notice that the ring breakings or
crackings generally occur around the places where the local kinetic energy $T_{m_{F}}$ is
vanishing, as compared with the (7,8) rows in Fig.~\ref{fig-1}. In fact, a large $T_{m_{F}}$ would enhance
the effect of the momentum-dependent SOC which actually exchanges the spin components. Otherwise, a vanishing $T_{m_{F}}$ might
weaken such a component exchange so that the density is not transferred, e.g., from $m_{F}=\pm 1$ to $m_{F}=0$,
thus leaving the vanishing density (cracking places) in the $m_{F}=0$ distribution.

\subsection{Phase distributions and symmetry breaking}

The phase distribution of $m_{F}=1$ [row (4)] in Fig.~\ref{fig-1} evolves from homogenous
structure (a), shape of two vertical long breads (b), anti-symmetric bow
ties in angle brackets (c), bow ties in distorted braces (d), vertical
stripes (e), eccentric rings (f), to horizontal stripes (g).

During these variations the rotational
symmetry in (a) in the absence of SOC is broken into reflection symmetry
(antisymmetric) in the $y$ ($x)$ direction in (b-d) in the presence of SOC,
the $y$-reflection symmetry is further broken in (f) as the rings are
eccentric. The $y$-reflection symmetry tend to recover in (g) but in an
antisymmetric way. The $m_{F}=0$ component [row (5)] has parity symmetry, while the
$m_{F}=\pm 1$ components are antisymmetric in parity within themselves and
symmetric under spin exchange.

\begin{figure}[t]
	\includegraphics[width=1.0\columnwidth]{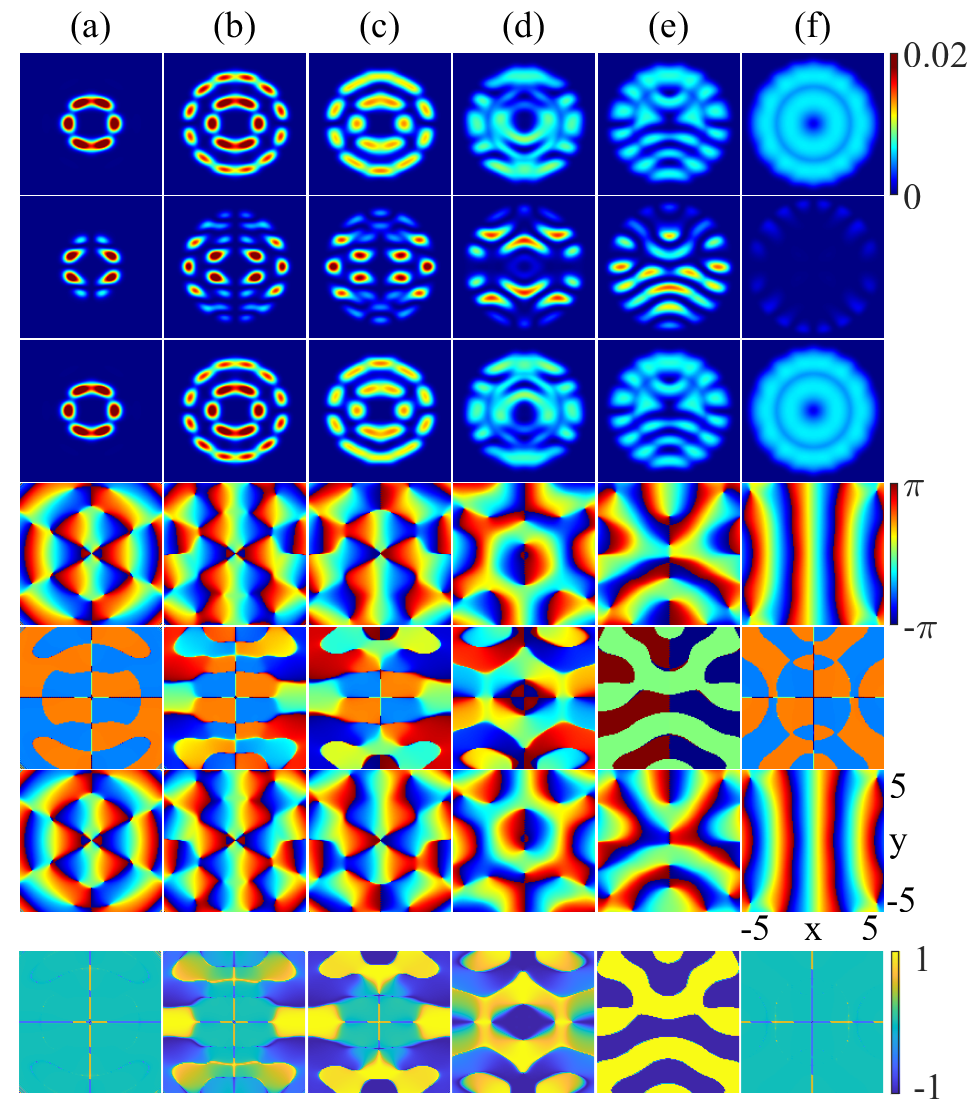}
	\caption{Effect of the interaction ratio on the GS structures at fixed SOC $\kappa=2$. (a) $\gamma=0.1$, (b) $\gamma=1$, (c) $\gamma=5$, (d) $\gamma=20$, (e) $\gamma=30$, (f) $\gamma=50$, with $c_{2}=55$. The upper three rows and middle three rows denote the density and phase for $m_{F}=+1,0,-1$ respectively, while the bottom row shows cosine of the phase for $m_{F}=0$.}
	\label{fig-Interaction-Ratio}
\end{figure}

\subsection{A tendency of density-phase separation}

A closer comparison of the phase and density draws our attention to a peculiar phenomenon of density-phase separation that the variations of density phase tend to be
independent.  In Fig.~\ref{fig-1}(4-6) we have plotted the phase $\phi _{m_{F}}$ within the range of $[-\pi,\pi]$, while a spurious boundary of $2\pi$ phase jump may arise even if the phase continuously crosses $\pm\pi$. On the other hand, it would be also necessary to distinguish a nontrivial phase variation of odd times of $\pi$ from even times~\cite{Ying2020PRR,Ying2017curvedSC,Ying2016Ellipse,Nagasawa2013Rings}.  For these reasons we replot the phase by $\cos \phi _{m_{F}}$ in Fig.~\ref{fig-1-phase}, with $m_{F}=1$ in the first row and $m_{F}=0$ in the second row. From the cases (a)-(e) in the second row of Fig.~\ref{fig-1-phase} we see that the amplitude of $\phi _{0}$ remains flat within the ring trap range despite that the various patterns of density distributions are changing, which indicates some degree of independence between the variations of the density and phase.  Line plots in Figs.~\ref{fig-1-phase}(4a-4e) and \ref{fig-1-phase}(3a-3e) by a fixed value of $x$ confirm such a density-phase separation, as $\cos \phi _{m_{F}}$ basically keeps in a flat status while the density are varying. In the other direction by fixing a $y$ position we also see such a tendency of density-phase separation in Figs.~\ref{fig-1-phase}(6a-6e) for $m_{F}=0$. Although some phase variation comes up in Figs.~\ref{fig-1-phase}(5a-5e) for $m_{F}=1$ as well as in Figs.~\ref{fig-1-phase}(6f-6g) for $m_{F}=0$, the correlation with the density variation is still not obvious, as reflected by the unmatched peak numbers.

The density-phase separation is partially broken by a very strong SOC as in Figs.~\ref{fig-1-phase}(3g,3h) and ~\ref{fig-1-phase}(4g,4h) where phase jumps by odd times of $\pi$ emerge between the density peaks. Still, the phase amplitude within each peak is nearly flat.

\subsection{A difference in vertical and horizontal-stripe states}

We have seen two orientations of the stripe states in the phase of the $%
m_{F}=\pm 1$ components, with vertical stripes in \ref{fig-1}(e4) and
horizontal stripes in \ref{fig-1}(g4). Still, despite of the
stripe-structure similarity of the phase in the $m_{F}=\pm 1$, they are
distinguished in the density of all spin components and the phase of the $%
m_{F}=0$ component. The horizontal one forms a stripe structure in the
density as well, while the vertical one does not show any stripe sign in the
density distribution of all the spin components as in \ref{fig-1}(g), the
latter also being a manifest of density-phase separation. Furthermore, when
in the horizontal one the phase of the $m_{F}=0$ component follows the
stripe structure of the $m_{F}=\pm 1$ components, the vertical one remains
in a flat amplitude of phase in the $m_{F}=0$ component as in~\ref%
{fig-1-phase}(e) and \ref{fig-Interaction-Ratio}(f). Such a difference
demonstrates the different roles of the unconventional SOC in the different
directions.

\subsection{Influence of interactions}

In previous sections we have addressed the effect of the SOC in inducing the exotic quantum states, here we examine the influence of the interactions. Fig.~\ref{fig-Interaction-Ratio} presents the variations of the density (upper three rows) and phase (lower three rows) distributions in varying the ratio between the density interaction and spin interaction $\gamma = c_0/ c_2$ at a fixed SOC. When we vary $\gamma$ from a small value [$\gamma =0.1$ in Fig.~\ref{fig-Interaction-Ratio}(a)] to a large one [$\gamma =50$ in Fig.~\ref{fig-Interaction-Ratio}(f)] the density distribution changes from  structures of a single broken inner ring (a), inner and outer broken rings (b,c), overlapping inner and outer broken rings (d), coexisting petals and stripes (e), to be a structure of two continuous rings in the $m_{F}=\pm 1$ components with vanishing inner ring and a weak outer fragmented ring in the $m_{F}=0$ component. Such variations are reasonable as the strengthened density-density interaction tends to extend the density distribution to reduce its positive energy, while the interplay of the interactions with the SOC induces the nontrivial variations.

We also see that the reflection symmetry with respect to the $x$ axis within a same component is basically preserved in small values of $\gamma$ as in Fig.~\ref{fig-Interaction-Ratio}(a-c) while it is broken in the $m_{F}=\pm 1$ components in Fig.~\ref{fig-Interaction-Ratio}(d) and in all spin component in Fig.~\ref{fig-Interaction-Ratio}(e). The symmetry tends to recover when the density turns to unbroken rings in larger $\gamma$ as in Fig.~\ref{fig-Interaction-Ratio}(f). The phase is more sensitive to check the symmetry breaking, as we can see from Fig.~\ref{fig-Interaction-Ratio}(c) the phase reflection anti-symmetry in the $m_{F}=\pm 1$ components is already broken obviously even when the density still basically preserves the reflection symmetry. Although in Fig.~\ref{fig-Interaction-Ratio}(c) and \ref{fig-Interaction-Ratio}(d) the anti-reflection symmetry is broken within a same component for $m_{F}=\pm 1$, it is holding if the spin is reversed simultaneously. Such a spin-exchange-associated reflection is also broken in Fig.~\ref{fig-Interaction-Ratio}(e) both in the desnity and phase.

Furthermore, we recall the afore-mentioned tendency of density-phase separation in the small-$\gamma$ case [Fig.~\ref{fig-Interaction-Ratio}(a)] and large-$\gamma$ case [Fig.~\ref{fig-Interaction-Ratio}(f)] as shown by the cosine plots of the phase for the $m_{F}=0$ component in the bottom row. We indeed see again here that the phase amplitude becomes flat when the density has a non-trivially inhomogeneous distribution.

\section{Ferromagnetic case}\label{Sect-FM}

In this part, we discuss the ferromagnetic case with $c_{2}<0$. Differently from the antiferromagnetic case, here we will see semi-circular or half-disk status of density
embedded with vortices and anti-vortices. We also find other exotic quantum states including self-arranging array of
half-skyrmions and half-antiskyrmion fence in spin distribution.

\begin{figure}[t]
    \includegraphics[width=1.0\columnwidth]{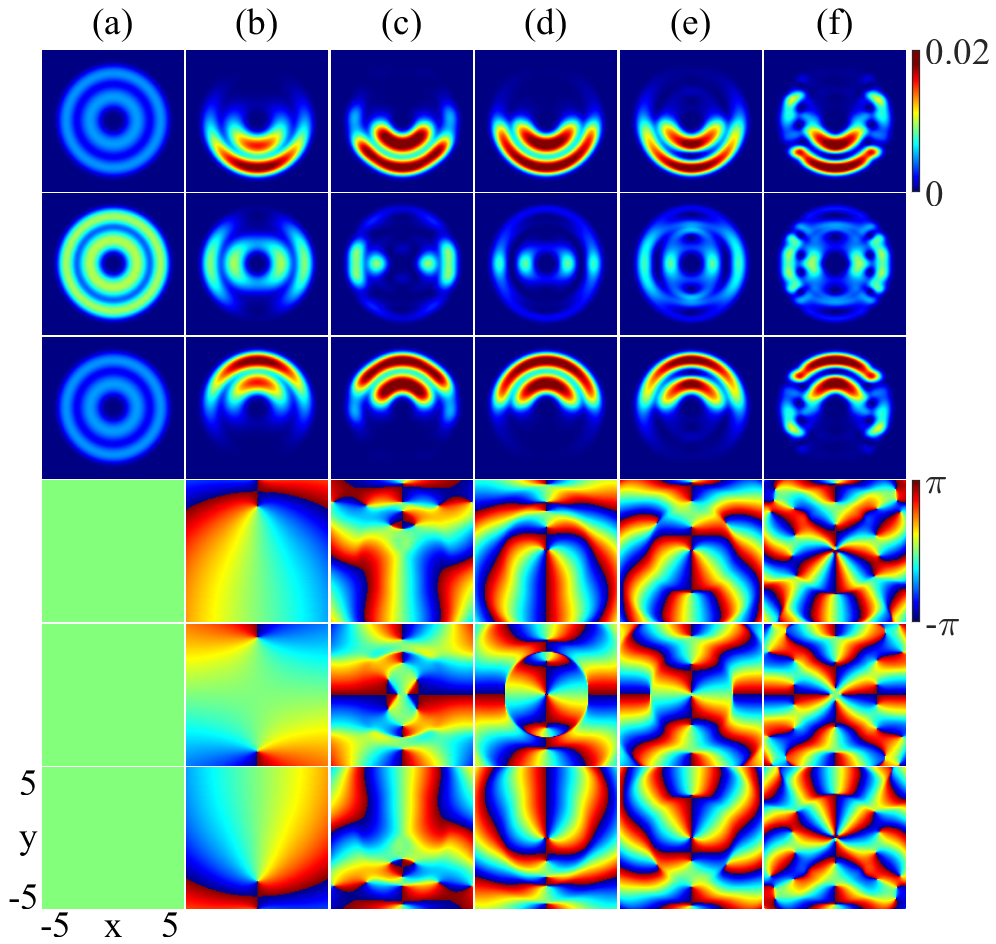}
	\caption{GS distributions of density (upper three rows) and phase (lower three rows) for $m_{F}=1, 0,-1$ in antiferromagnetic interaction $c_0=550$ and $c_2=-55$ for different SOC $\kappa$. (a) $\kappa=0$, (b) $\kappa=0.4$, (c) $\kappa=0.8$, (d) $\kappa=1.2$, (e) $\kappa=1.6$, (f) $\kappa=2$, (g) $\kappa=2.5$. }
	\label{fig-antiF-vary-SOC}
\end{figure}
\begin{figure}[t]
    \includegraphics[width=1.0\columnwidth]{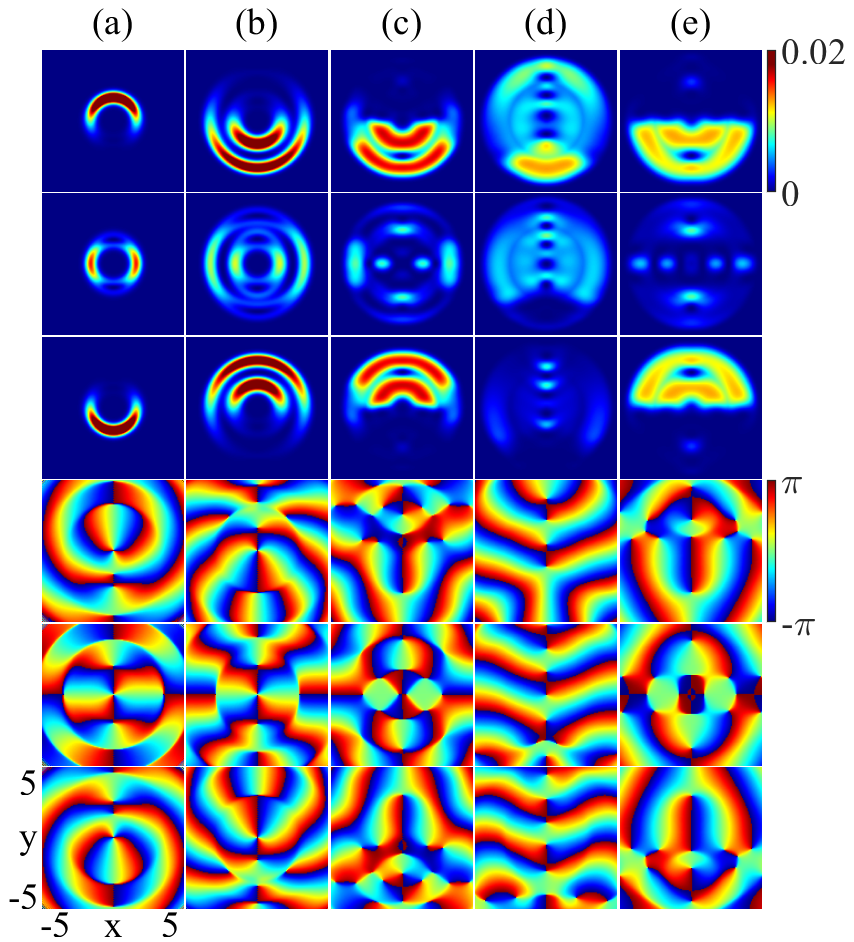}
	\caption{Effect of the interaction ratio $\gamma$ on the GS distributions of density (upper three rows) and phase (lower three rows) for $m_{F}=1, 0,-1$. at fixed SOC $\kappa=2.0$. (a) $\gamma=-1$, (b) $\gamma=-5$, (c) $\gamma=-20$, (d) $\gamma=-50$, (e) $\gamma=-50$, $\kappa=2.0$ with $c_{2}=-55$. $\kappa=2.0$ for (a)-(d) and $\kappa=1.6$ for (e).}
	\label{fig-antiF-vary-Interaction}
\end{figure}

\subsection{Semi-circular distribution in varying SOC}

We first check the GS structure in varying the SOC at fixed
density-density and spin-spin interactions.

In the absence of SOC, the
density appears with a structure of two concentric rings in all components,
as in Fig.~\ref{fig-antiF-vary-SOC}(a). Here, in contrast to the vanishing
density in the antiferromagnetic case, the $m_{F}=0$ component has a
finite density and the density is even larger than the $m_{F}=\pm 1$
components.

Turning on the SOC, in the $m_{F}=0$ component the density tends
to gather around the $x$ axis, while away from the $x$ axis the density is
smaller, as in Figs.~\ref{fig-antiF-vary-SOC}(b) and \ref{fig-antiF-vary-SOC}(c).
With an increased strength of SOC in Figs.~\ref{fig-antiF-vary-SOC}(c)
and \ref{fig-antiF-vary-SOC}(d), we see that fissures emerge in the weak parts of
density rings around the $y$ axis. In a larger SOC, the outer and inter rings
tend to get connected, and the fissures evolve into holes as in Fig.~\ref{fig-antiF-vary-SOC}(e).
With further region connection more density holes
form also away from the $y$ axis, as in Fig.~\ref{fig-antiF-vary-SOC}(f).

On the
other hand, in the $m_{F}=\pm 1$ components, the density distributions become
half-ring-like in the presence of SOC, with lower half rings for $m_{F}=+1$ and
upper half rings for $m_{F}=-1$, as in Figs.~\ref{fig-antiF-vary-SOC}(b)-(e).
In a large SOC, the density holes emerge too,
also vertically but displaced from those of $m_{F}=0$ components, together
with the remaining density arches forming a clown-face like distribution.

We also plot the phase in the lower three rows of Fig.~\ref{fig-antiF-vary-SOC}.
The phase evolves from a uniform distribution in the
the absence of SOC into a complicated distribution even when the density
basically remains in a half-ring structure in the $m_{F}=\pm 1$ components,
which also manifests a density-phase separation in some sense.

\subsection{Influence of interaction ratio}

The variation of the GS in different interaction ratio $\gamma $ is
presented in Fig.~\ref{fig-antiF-vary-Interaction}. With the increase of
$\gamma $, the density distribution in $m_{F}=\pm 1$ changes the shape from an
inner half ring (a), inner and outer half rings (b), inner and outer arcs
(c), to coexisting arcs and half disk with holes (d). We also find a state of half disk with dual
hole and island (e) by tuning a bit the SOC at the final interaction ratio, which more confirms the manipulation effect by the interplay of the SOC and interactions.

In the $m_{F}=0$ component, the density distribution
varies from inner structure with left-right arcs and top-bottom fissures
(a), inner and outer structures with left-right arcs and top-bottom fissures
(b), astronaut-helmet-like shape (c), corner-broken disk (d), to coexisting
half disks and island array (e).

\begin{figure}[t]
    \includegraphics[width=1.0\columnwidth]{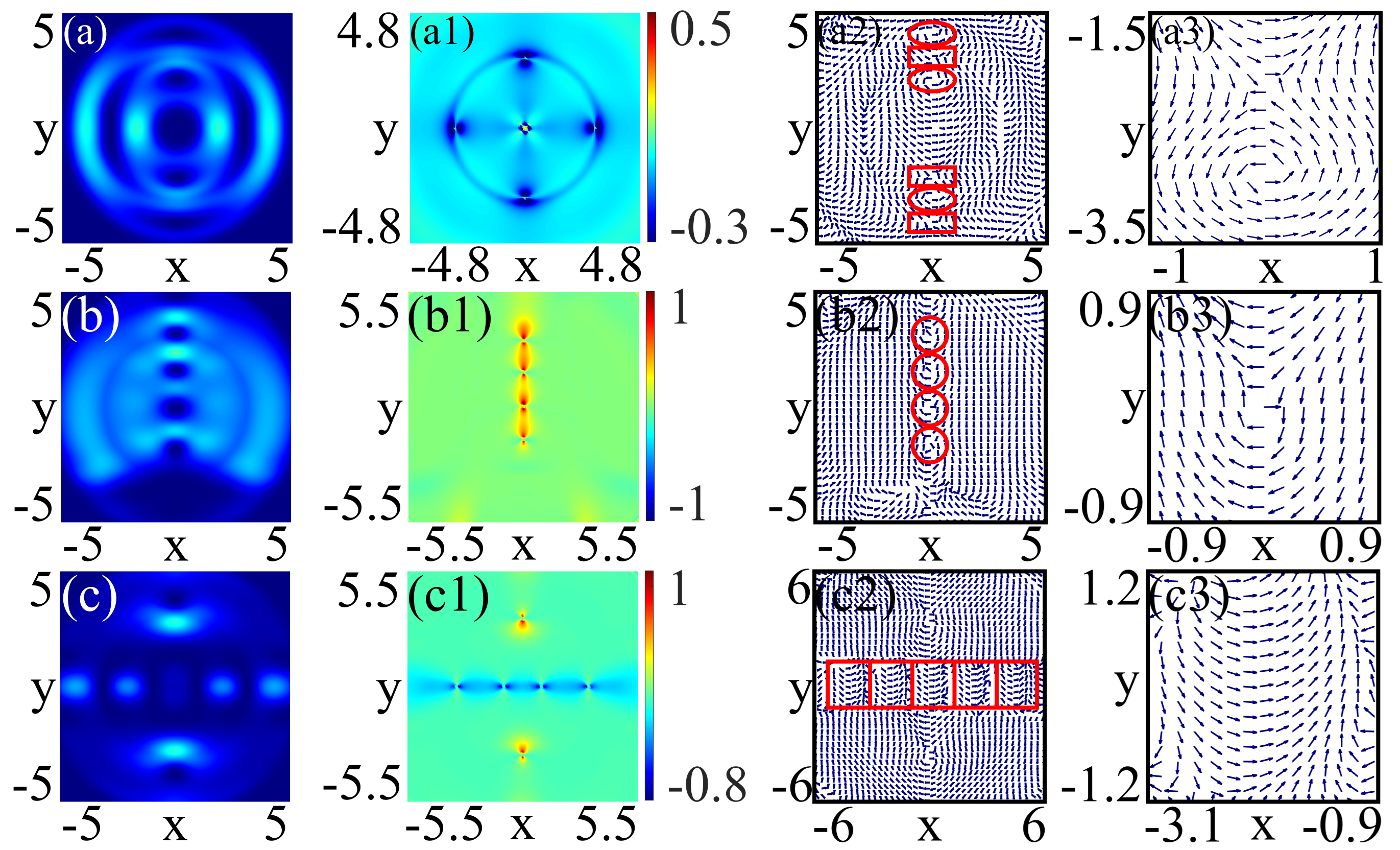}
	\caption{Different states of topological defects. (a,b,c) Density of $m_F=0$ component, (a1,b1,c1) topological charge density,  (a2,b2,c2) in-plane spin distribution,  (a3,b3,c3) close-up view of the topological defects, with parameters:
(a)-(a3) $\kappa=2.0$, $\gamma=-10$, $c_2=-55$;
(b)-(b3) $\kappa=2.0$, $\gamma=-50$, $c_{2}=-55$;
(c)-(c1) $\kappa=1.6$, $\gamma=-50$, $c_{2}=-55$.
}
	\label{fig-Votex-antiVortex-3}
\end{figure}

\subsection{Topological defects in spin distribution}\label{Sect-topo-defect}

The spin texture can be extracted by~\cite{Mizushima2004defSpin1Texture,Kasamatsu2005defSpin1Texture}
\begin{align}
	S_{\alpha}=\sum_{m_F,m^\prime_F=0,\pm1}\psi_{m_F}^{\ast}(\widehat{F}_{\alpha})_{m_F,m^\prime_F}\psi_{m^\prime_F}/|\psi|^{2},
\end{align}
where $\alpha=x,y,z$ label the three spin directions. Topological defects may arise in the spin texture in our system. The spatial structure of the topology of the system can be
described by the topological charge density
\begin{equation}
q(\bm r)=\frac{1}{4\pi }\bm s\bm{\cdot}\left( \frac{\partial \bm s}{\partial
x}\times \frac{\partial \bm s}{\partial y}\right)   \label{q-topo-density}
\end{equation}%
where $\bm s=\bm S/|\bm S|$ and $\bm S=\{S_x,S_y,S_z\}$, while its integral
\begin{equation}
Q=\int q(\bm r)dxdy.  \label{Q-topo-charge}
\end{equation}
characterizes the topological charge~\cite{Nagaosa2013Skyrmions}.

\subsubsection{Vortices and antivortices}\label{Sect-Vortx-antiVortx}

We find that vortices and antivortices emerge in all the states with
half-ring structure. The case with a weak SOC in Fig.~\ref{fig-antiF-vary-SOC}(b)
has one vortex and one antivortex,
while their number increases when the $m_{F}=0$ component has fissures and
holes in the density as the SOC is enhanced. Indeed the vortices and antivortices
form at the density valleys and peaks of the $m_{F}=0$ component. In Fig.~\ref{fig-Votex-antiVortex-3}(a2),
we illustrate the spin projection on the $x $-$y$ plane for the case in Fig.~\ref{fig-antiF-vary-Interaction}(b) [similar to Fig.~\ref{fig-antiF-vary-SOC}(e)].
The vortices
are marked by the ellipses and antivortices by rectangles in Fig.~\ref{fig-Votex-antiVortex-3}(a2)
and the structure of the vortex and antivortex can
be seen more clearly in a close-up plot in Fig.~\ref{fig-Votex-antiVortex-3}(a3).

We notice two features about the vortex locations in our system: (i) These vortices and antivortices are distributed
antisymmetrically with respect to the $x$ axis in the sense that if a vortex
appears at $y$ position an antivortex will form at position $-y$. (ii)
These vortices and antivortices are located at the position density valleys
(holes) and peaks of the $m_{F}=0$ component, as compared with the density
plot in Fig.~\ref{fig-Votex-antiVortex-3}(a). It should be noted that the valleys
may correspond to density peaks in the $m_{F}=\pm 1$ components due to the
fissure or hole dislocations in reducing the energy of the density-density
interaction.

It should be mentioned that one can also find vortices and
antivortices in the ring-fissure state [Fig.\ref{fig-1}(f)] and the stripe state [Fig.\ref{fig-1}(g)] in the antiferromagnetic case addressed
in Section~\ref{Sect-AntiF}, with a similar location tendency as in the above
features. Indeed, the ring-fissure state has a series of vortices (antivortices) on the positive-(negative-) $y$ axis,
while in the stripe state more vortices and antivortices are dispersedly distributed at the wavy places of the stripes.

These observations indicate that the locations of vortices and
antivortices can be manipulated by tuning the potential geometry, SOC and interactions.

\subsubsection{Self-aligning half-skyrmion array}

The spin texture can also self-arranges into an array of half-skyrmions. We
illusrate such a state in Fig.~\ref{fig-Votex-antiVortex-3}(b2) where the
half-skyrmion array is marked by the circles. Such a half-skymion alignment
is associated with the hole arrray in the density disk in Fig.~\ref{fig-antiF-vary-Interaction}(d)
with the $m_{F}=0$ component replotted in
Fig.~\ref{fig-Votex-antiVortex-3}(b) for better position correspondence with the half-skyrmions. A close-up view of the half-skyrmion spin
structure is presented in Fig. \ref{fig-Votex-antiVortex-3}(b3).

The topological charge density $q(\bm r)$ is shown in Fig.~\ref{fig-Votex-antiVortex-3}(b1). The
half-skyrmion has a fractional topological charge around
\begin{equation}
Q \sim 0.5
\end{equation}
as defined by \eqref{Q-topo-charge}.
Indeed we have a local topological charge $Q = 0.498$ by numerical integral of $q(\bm r)$ over the local area around a half-skyrmion illustrated in Fig.~\ref{fig-Votex-antiVortex-3}(b3).

\subsubsection{Self-arranging half-antiskyrmion fence}

We also find a half-antiskyrmion fence that forms horizontally along the
$x$-axis and separates two vortix-antivortex pairs located at the positive and
negative $y$-axes, as shown in Fig.~\ref{fig-Votex-antiVortex-3}(c2) where each square marks a half-antiskyrmion.
The half-antiskyrmions have a minus fractional topological charge
\begin{equation}
Q\sim-0.5.
\end{equation}
Such a half-antiskyrmion fence is associated with the density distribution
of coexisting half disks and island array in Fig.~\ref{fig-antiF-vary-Interaction}(e), as compared
with density in the $m_{F}=0$ component replotted in
Fig.~\ref{fig-Votex-antiVortex-3}(c) for better position correspondence. The corresponding topological charge
density is given in Fig.~\ref{fig-Votex-antiVortex-3}(c1) while a zoom-in
plot of a half-antiskyrmion in the fence is displayed in Fig.~\ref{fig-Votex-antiVortex-3}(c2).

\section{Conclusions and discussions}

\label{Conclusion}

We have systematically investigated a spin-1 BEC with an unconventional SOC
in two concentric annular traps, the GS of which
manifests various exotic quantum states from the distributions of the
density, phase and spin texture.

In the antiferromagnetic case, the GS density exhibits various patterns of
distributions, including ring state, petal states, facial-makeup states,
topological fissure states, multiple-half-ring states, property-distinguished vertical and horizonal stripe states.
In particular, the fissures in the ring density not only bring changes in the geometric topology of the density but also are
topologically associated with appearance of spin vortices and antivortices.
In the strength increase of the SOC, the symmetry is broken in different
degrees, with respect to the rotation symmetry, reflection symmetry, parity
symmetry, within a same spin component or simultaneously with a spin
exchange. The cracking places in density ring fragmentation are related with
vanishing local kinetic energy which locally weakens the effect of the SOC. We also notice a peculiar
phenomenon of density-phase separation in the sense that the variations of
density and phase tend to be independent.

In ferromagnetic case, the GS displays density distributions different from the antiferromagnetic case by
semi-circular structure, corner-broken disk with aligned holes, or
coexisting half disk and island array. The spin distribution is embedded
with vortices and anti-vortices. The spin distribution can self-arranges
into an array of half-skyrmions aligning vertically. A horizontal half-antiskyrmion
fence is also found separating vortex-antivortex pairs.

The topological defects generally appear around the density valleys and
peaks. In antiferromagnetic case vortices can form on the $y$ axis in ring-fissure states and more dispersedly in wavy places of stripe states,
while in ferromagnetic case the topological
defects are located on the $x$ or $y$ axis.   Indeed, vortices and
anti-vortices always can be found on the $y$ axis in semicircular states in
ferromagnetic case. Half-skyrmion array along the $y$ axis and
half-antiskyrmion fence along the $x$ axis emerge in the strong overlapping
of density rings, as driven by a large ratio of density interaction
and spin interaction in the presence of the SOC. These observations
indicate that the unconventional SOC may be helpful to create stripe states
and locate the positions of topological defects.

These exotic quantum states in the various distribution patterns in the
density, phase and spin texture with topological structures are the results
of the interplay of the SOC, ratio of density interaction and
spin interaction and the geometry of potential. Our study implies
that one can manipulate the emergence of exotic quantum states via these
factors, which is feasible due to the high controllability of cold atom
systems~\cite{Li2012PRL,LinRashbaBECExp2011,LinRashbaBECExp2013Review,wu2016realizeSOC,Campbell2016RealizeSpin1SOC,GongMing2019BECSOC,Anderson2013PRLmagnGenerateSOC}. As a final remark, the abundant state variations might also provide potential
resources for quantum metrology~\cite{RamsPRX2018,Garbe2020,Montenegro2021-Metrology,Ilias2022-Metrology,Ying2022-Metrology,Hotter2024-Metrology,
Ying-Topo-JC-nonHermitian-Fisher,Ying-g2hz-QFI-2024} by exploiting the change of the GS wave
function for measurement of parameters~\cite{Cramer-Rao-bound,RamsPRX2018,Ying-gC-by-QFI-2024}.

\bigskip

\begin{acknowledgments}
	
This work was supported by the National Natural Science Foundation of China
(Grants No. 12474358, No. 11974151, and No. 12247101).

\end{acknowledgments}

\bibliography{ref}

\end{document}